\documentclass[sigplan]{acmart}

\setcopyright{acmlicensed}
\acmPrice{15.00}
\acmDOI{10.1145/3372885.3373824}
\acmYear{2020}
\copyrightyear{2020}
\acmISBN{978-1-4503-7097-4/20/01}
\acmConference[CPP '20]{Proceedings of the 9th ACM SIGPLAN International Conference on Certified Programs and Proofs}{January 20--21, 2020}{New Orleans, LA, USA}
\acmBooktitle{Proceedings of the 9th ACM SIGPLAN International Conference on Certified Programs and Proofs (CPP '20), January 20--21, 2020, New Orleans, LA, USA}

\usepackage[utf8x]{inputenc}
\usepackage[T1]{fontenc}
\usepackage{graphicx}
\usepackage{grffile}
\usepackage{longtable}
\usepackage{supertabular}
\usepackage{wrapfig}
\usepackage{rotating}
\usepackage[normalem]{ulem}
\usepackage{amsmath}
\usepackage{textcomp}
\usepackage{amssymb}
\usepackage{capt-of}
\usepackage{scalefnt}
\usepackage{color}
\definecolor{keywordcolor}{rgb}{0.7, 0.1, 0.1}   
\definecolor{tacticcolor}{rgb}{0.1, 0.2, 0.6}    
\definecolor{commentcolor}{rgb}{0.4, 0.4, 0.4}   
\definecolor{symbolcolor}{rgb}{0.0, 0.1, 0.6}    
\definecolor{sortcolor}{rgb}{0.1, 0.5, 0.1}      
\definecolor{attributecolor}{rgb}{0.7, 0.1, 0.1} 
\usepackage{listings}

\usepackage{xspace}
\usepackage{tikz}
\usepackage{ellipsis}
\usepackage{flushend}

\newcommand{\mathlib}{\textsf{mathlib}\xspace}
\newcommand{\corelib}{\textsf{core} library\xspace}
\newcommand{\mathcomp}{Mathematical Components\xspace}
\newcommand{\numcontrib}{73\xspace}
\newcommand\smul[1][.5]{\mathbin{\vcenter{\hbox{\scalebox{#1}{$\bullet$}}}}}

\begin{document}

\title
      {The Lean Mathematical Library}


\author{The \mathlib Community}
\authornote{This paper describes a community effort with \numcontrib contributors, listed in the acknowledgments. Rather than attributing authorship to a subset, we have decided to release this paper under a collective pseudonym.} 
\affiliation{
}


\begin{abstract}
This paper describes \mathlib,
a community-driven effort to build a unified library of mathematics
formalized in the Lean proof assistant.
Among proof assistant libraries,
it is distinguished by
its dependently typed foundations,
focus on classical mathematics,
extensive hierarchy of structures,
use of large- and small-scale automation,
and distributed organization.
We explain the architecture and design decisions of the library
and the social organization that has led to its development.
\end{abstract}



\maketitle


\section{Introduction}
\label{section:intro}

Since the first mechanized proof-checking systems were released, there has been continual effort to develop libraries of formal definitions and proofs.
Every major system has at least one library that can serve as a base for further formalizations.
The contents, organizations, and purposes of these libraries vary considerably.
Some are small, static cores bundled with the system itself;
some are sprawling repositories that solicit contributions like a journal; some focus on a narrow subject area, while others are more broad-minded in their topics.

This paper describes \mathlib, a formal library developed for the Lean proof assistant \cite{Mour15a}.
As a community-driven effort with dozens of contributors, there is no central organization to \mathlib;
it has arisen from the desires of its users to develop a repository of formal mathematical proofs.
We are certainly not the first to profess this goal~\cite{QED94}, nor is our library particularly large in comparison to others.
However, its organizational structure,
focus on classical mathematics,
and inclusion of automation
distinguish it in the space of proof assistant libraries.
We aim here to explain our design decisions and the ways in which \mathlib has been put to use.

In contrast to most modern proof assistant libraries, many of the contributors to \mathlib have an academic background in pure mathematics.
This has significantly influenced the contents and direction of the library.
It is a goal of many in the community to support the formalization of modern, research-level mathematics, and various projects discussed in Section~\ref{subsection:community:oneoff} suggest that we are approaching this point.

\subsection{A History of \mathlib and Lean 3}
\label{subsection:intro:history}

The Lean project was started by Leonardo de Moura in 2013~\cite{Mour15a}.
Its most recent version, Lean 3, was released in early 2017~\cite{EURAM17}.
A new version is under development~\cite{Ullr19}.
In the summer of 2017, much of the core library for Lean was factored out of the system repository.
This code became the base for \mathlib.
The project was initially led by Mario Carneiro and Johannes H\"olzl at Carnegie Mellon University,
and attracted a growing number of users,
who were drawn by the system documentation~\cite{Avig14}, the library's focus on classical mathematics, and the real-time chat on Zulip.

Over the two years since the split, \mathlib has grown from 15k to 140k lines of code (LOC), excluding blank lines and comments. 
Contributions have been made by \numcontrib people and are managed by a team of 11 maintainers.
It is the de facto standard library for both programming and proving in Lean 3.
The surrounding community has developed infrastructure, promotional materials, and university courses based on \mathlib.

\section{Lean}
\label{section:lean}

Like Coq, Lean uses a system of dependent types based on the calculus of inductive constructions (CIC)~\cite{Pfen89}.
Lean has a noncumulative hierarchy of universes
\lstinline{Prop}, \lstinline{Type}, \lstinline{Type 1}, \lstinline{Type 2}, \lstinline{Type 3} \ldots\,
The bottom universe \lstinline{Prop} is \emph{impredicative}, meaning that quantification of a \lstinline{Prop} over a larger type is a \lstinline{Prop}, and \emph{proof-irrelevant}, meaning that all proofs of the same proposition are definitionally (or judgmentally) equal.

Lean adds two axioms to the type theory.
The first axiom, \lstinline{classical.choice}, produces a term of type \lstinline{T} from a (propositional) proof that \lstinline{T} is nonempty.
This is a type-theoretic statement that implies the set-theoretic axiom of choice.
The second axiom, \lstinline{propext}, states that a bi-implication between two propositions implies that these propositions are equal.

Lean also extends the CIC with \emph{quotient types}.
Given an equivalence relation on a type, the quotient type identifies related terms.
This feature allows us to work with equality of objects in the quotient
and removes the need for setoids.

A key feature of Lean 3 is its \emph{metaprogramming} framework~\cite{EURAM17}.
This allows users to write tactics, commands, and other tools for manipulating the Lean environment directly in the language of Lean itself.
In \mathlib, we make extensive use of metaprogramming both for powerful automation and specialized commands to ease various tasks when formalizing.
We will discuss this more in Section~\ref{section:metaprogramming}.

\section{Contents of \mathlib}
\label{section:contents}

The \mathlib library is designed as a basis for research level mathematics, as well as a standard library for programming in Lean.
We build \mathlib on top of a small \corelib, shipped with Lean, which contains 19k LOC.
The \corelib sets up the meta\-programming and tactic framework for Lean, but it also contains much of the algebraic hierarchy and the definitions of basic datatypes, like $\mathbb{N}$, $\mathbb{Z}$, and \lstinline{list}.

Currently, much of \mathlib consists of undergraduate level mathematics.
Table~\ref{table:directories} gives the size of all top-level directories in the \texttt{src} directory.
This gives a rough idea of the relative sizes of the different topics in \mathlib, but the contents of some directories require some extra explanation.

\begin{table}
  \begin{tabular}{lrr}
    \bf Subdirectory & \bf LOC & \bf Declarations  \\ \hline
    \texttt{data} & 41849 & 10695 \\
    \texttt{topology} & 17382 & 2709 \\
    \texttt{tactic} & 12184 & 1679 \\
    \texttt{algebra} & 9830 & 2794 \\
    \texttt{analysis} & 7962 & 1237 \\
    \texttt{order} & 6526 & 1542 \\
    \texttt{category\_theory} & 6299 & 1560 \\
    \texttt{set\_theory} & 6163 & 1394 \\
    \texttt{measure\_theory} & 6113 & 926 \\
    \texttt{ring\_theory} & 5683 & 1080 \\
    \texttt{linear\_algebra} & 4511 & 805 \\
    \texttt{computability} & 4205 & 575 \\
    \texttt{group\_theory} & 4191 & 1094 \\
    \texttt{category} & 1770 & 389 \\
    \texttt{number\_theory} & 1394 & 228 \\
    \texttt{logic} & 1195 & 403 \\
    \texttt{field\_theory} & 1002 & 121 \\
    \texttt{geometry} & 848 & 70 \\
    \texttt{meta} & 784 & 135 \\
    \texttt{algebraic\_geometry} & 194 & 29 \\ \hline
      & 140085 & 34168
  \end{tabular}
  \caption{Lines of code, excluding white space and comments, in the top-level directories of the \mathlib source code as of December 12, 2019.}
  \label{table:directories}
\end{table}

In the \texttt{algebra} library, we expand on the \corelib algebraic hierarchy, from \lstinline{semigroup} to \lstinline{linear_ordered_field}, \lstinline{module}, and more (Section~\ref{subsection:typeclasses:hierarchy}).
The results in this folder are focused on computing with elements in these algebraic structures.
The structural theory of these algebraic objects is developed in the folders \texttt{group\_theory}, \texttt{ring\_theory}, \texttt{field\_theory}, and \texttt{linear\_algebra}.

The \texttt{data} folder contains the definitions and properties of data structures,
including the number systems $\mathbb{Q}$, $\mathbb{R}$, and $\mathbb{C}$,
sets and subtypes,
partial and finitely supported functions,
polynomials, lists, multisets, and vectors.

The folders \texttt{meta} and \texttt{tactic} use Lean's metaprogramming framework (Section~\ref{section:metaprogramming}) to define custom tactics. The \texttt{category} folder defines infrastructure for categorical programming, as used in Haskell; this is not to be confused with the \texttt{category\_theory} folder, which develops the mathematical theory.

Quotients are heavily used in \mathlib to avoid the need for setoids.
They are used to define multisets as lists up to permutation, which are in turn used to define finite sets as multisets without duplicates.
Quotients are also frequently used in algebra, for example for the definition of a quotient group, the tensor product of modules or the colimit of rings.
We also use quotients when defining the Stone-\v Cech compactification and Cauchy completion, the latter of which is used to define the real and $p$-adic numbers.
Quotients are also used to define cardinals (as types modulo equivalence) and ordinals (as well-ordered types modulo order-isomorphism).


The extensive \texttt{topology} library includes theories about uniform spaces, metric spaces,
and algebraic topological spaces such as topological groups and rings.
A more novel feature is the definition of the Gromov--Hausdorff space,
i.e.\ the space of all nonempty compact metric spaces up to isometry,
with its natural (Gromov--Hausdorff) distance,
and the proof that it is a Polish space.\footnote{
The possibility of formalizing the Gromov--Hausdorff space
was the motivation for the switch of a user from Isabelle/HOL to Lean,
as it makes heavy use of the machinery of dependent types.}

The definition of a manifold in mathlib is very general compared to other known formalizations \cite{DBLP:conf/cpp/ImmlerZ19,Pak16}.
In particular,
the base field for differentiable manifolds can be any non-discrete normed field,
including $\mathbb{R}$, $\mathbb{C}$, and $\mathbb{Q}_p$.
Arbitrary models and structure groupoids can be used.
Examples currently include $C^k$ and $C^\infty$ manifolds,
possibly with boundary or corners.
Potential future examples include analytic manifolds,
contact or symplectic manifolds, and translation surfaces.

The \texttt{category theory} library includes (co)limits, monadic adjunctions, and monoidal categories.
It uses extensive automation to hide easy proofs (Section~\ref{subsection:metaprogramming:tactics}).
The theory established here is used in other parts of the library,
for example describing the Giry monad in measure theory,
or showing that products and equalizers in complete separated uniform spaces can be calculated in the underlying uniform spaces.

Since Lean's type theory contains a universe hierarchy, it has a higher consistency strength than ZFC, and the \texttt{set theory} library defines a model of ZFC. Additionally, cardinals and ordinals are defined using quotients of a universe, along with related notions like ordinal arithmetic, $\aleph_\alpha$, cofinality, inaccessible cardinals, etc. This machinery is used in other parts of the library~(Section~\ref{subsection:linalg:dimension}).

Structural results about groups, rings, and fields include Sylow's theorems, the law of quadratic reciprocity, integral and perfect closures, and Hilbert's basis theorem. The \texttt{linear algebra} library is described in detail in Section~\ref{section:linalg}.

A library on computability theory~\cite{Carn19} proves the undecidability of the halting problem,
and a library on the $p$-adic numbers proves Hensel’s lemma~\cite{Lewi19}.

Analysis is a weaker point of \mathlib, although the Fr\'echet derivative and the Bochner integral have been formalized, as well as basic properties of trigonometric functions.

\section{Type Classes}
\label{section:typeclasses}

Type classes are predicates or extra data attached to types, which are systematically inferred by a Prolog-like backtracking search.
The idea of using type classes to permit polymorphism over types with a particular structure was pioneered in Haskell~\cite{Wadl89}, and the usefulness of type classes for organizing mathematical structures and theorems on these structures was recognized in Isabelle~\cite{Wenzel97, Haft06} and Coq~\cite{Spit11}.
Lean's \corelib\ builds on these ideas by using type classes ubiquitously, including for:
\begin{itemize}
\item notations, e.g.\ the \lstinline{has_add α} type class that is inferred when the $+$ symbol is used on a type $\alpha$;
\item mathematical structures, e.g.\ the type \lstinline{ring α} that equips a type $\alpha$ with a ring structure;
\item coercions, via the type class \lstinline{has_coe α β}, a wrapper around the type \lstinline{α → β} that allows Lean to accept an element of $\alpha$ where a $\beta$ is expected; and
\item decidability of propositions, via the \lstinline{decidable p} type class, which enables case analysis in constructive contexts and if statements in programming.
\end{itemize}
This usage is significantly expanded into all parts of \mathlib.

Mathematicians and computer scientists often speak of an \emph{algebraic hierarchy}, but insofar as that brings to mind the groups, rings, and fields of an introductory course in abstract algebra, the phrase has the wrong connotations.
Lean does have a hierarchy of algebraic structures,
but this only scratches the surface.
For example, in \mathlib, the real numbers are an instance of a normed space, a metric space, a uniform space, and a normal topological space.
Morphisms between structures have algebraic properties that also need to be managed.
The library instantiates the category of groups and morphisms between them, and the functor mapping a measure space $X$ to the space of probability measures on $X$ is an instance of a monad, namely, the \emph{Giry monad}~\cite{Giry82}.
To convey the richness of such a network of definitions, we will refer to it rather as a \emph{structure hierarchy}.

\subsection{Organizing the Structure Hierarchy}
\label{subsection:typeclasses:hierarchy}

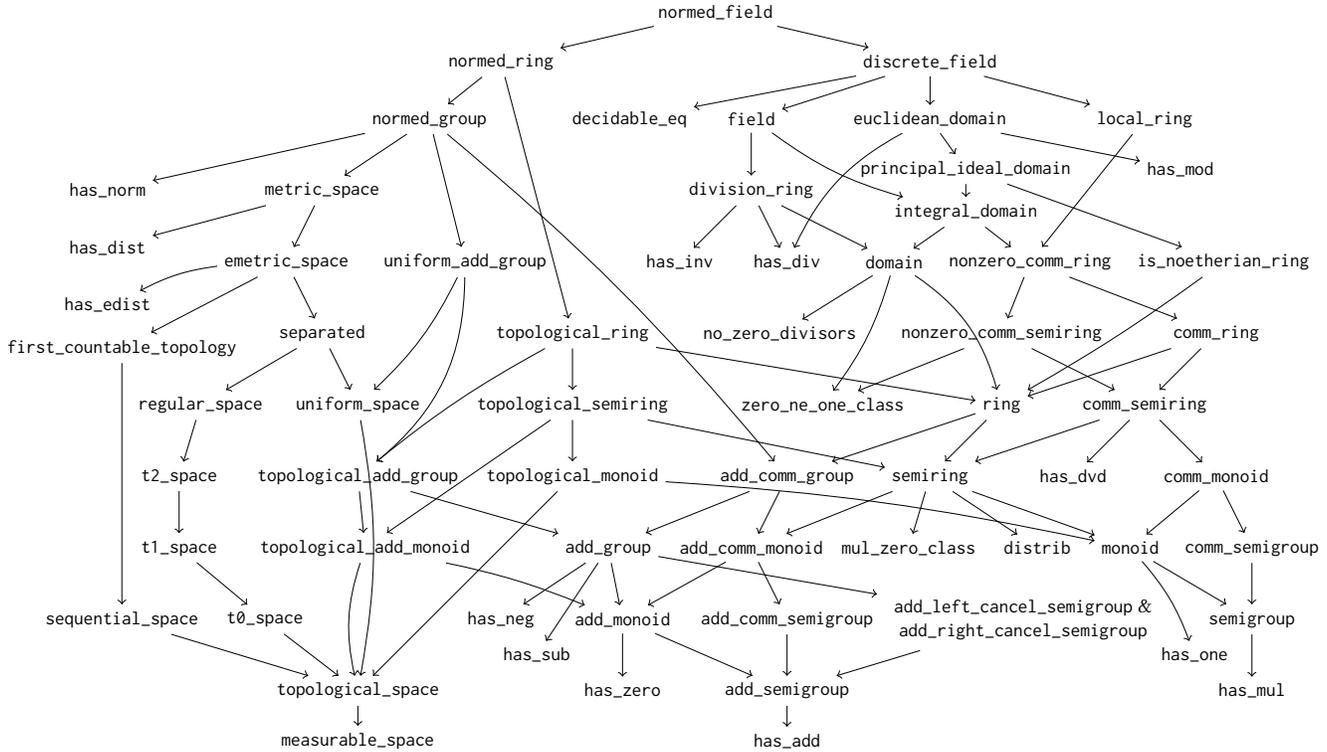
\begin{figure*}
 \tikzset{every node/.style={scale=0.8}}
 \resizebox{\textwidth}{!}{ 
 \begin{tikzpicture}[->]

\node (normed_field) at (1, 1.5) {\lstinline{normed_field}};

\node (normed_ring) at (-2, 0.8) {\lstinline{normed_ring}};
\node (discrete_field) at (4, 0.8) {\lstinline{discrete_field}};
\path (normed_field) edge (discrete_field);
\path (normed_field) edge (normed_ring);

\node (normed_group) at (-3, 0) {\lstinline{normed_group}};
\node (topological_ring) at (-1, -3) {\lstinline{topological_ring}};
\node (field) at (1.5, 0) {\lstinline{field}};
\node (decidable_eq) at (-0.2, 0) {\lstinline{decidable_eq}};
\node (euclidean_domain) at (4, 0) {\lstinline{euclidean_domain}};
\node (local_ring) at (7, 0) {\lstinline{local_ring}};
\path (discrete_field) edge (euclidean_domain);
\path (discrete_field) edge (field);
\path (discrete_field) edge (local_ring);
\path (discrete_field) edge (decidable_eq);
\path (normed_ring) edge (normed_group);
\path (normed_ring) edge (topological_ring);

\node (has_norm) at (-7.5, -1) {\lstinline{has_norm}};
\node (metric_space) at (-4.5, -1) {\lstinline{metric_space}};
\node (uniform_add_group) at (-2.5, -2) {\lstinline{uniform_add_group}};
\node (topological_semiring) at (-1, -4) {\lstinline{topological_semiring}};
\node (division_ring) at (1.5, -1) {\lstinline{division_ring}};
\node (integral_domain) at (4.5, -1.3) {\lstinline{integral_domain}};
\node (has_mod) at (7.5, -0.7) {\lstinline{has_mod}};
\node (principal_ideal_domain) at (4.5, -0.7) {\lstinline{principal_ideal_domain}};
\path (field) edge (division_ring);
\path (field) edge[bend right=10] (integral_domain);
\path (euclidean_domain) edge (has_mod);
\path (euclidean_domain) edge (principal_ideal_domain);
\path (normed_group) edge (uniform_add_group);
\path (normed_group) edge (metric_space);
\path (normed_group) edge (has_norm);
\path (topological_ring) edge (topological_semiring);

\node (topological_add_group) at (-4, -5) {\lstinline{topological_add_group}};
\node (emetric_space) at (-5, -2) {\lstinline{emetric_space}};
\node (has_dist) at (-7.5, -1.8) {\lstinline{has_dist}};
\node (topological_monoid) at (-1, -5) {\lstinline{topological_monoid}};
\node (has_inv) at (0.5, -2) {\lstinline{has_inv}};
\node (has_div) at (2, -2) {\lstinline{has_div}};
\node (domain) at (3.5, -2) {\lstinline{domain}};
\node (nonzero_comm_ring) at (5.4, -2) {\lstinline{nonzero_comm_ring}};
\node (is_noetherian_ring) at (8.1, -2) {\lstinline{is_noetherian_ring}};
\path (division_ring) edge (has_div);
\path (euclidean_domain) edge[bend right=18] (has_div);
\path (division_ring) edge (has_inv);
\path (division_ring) edge (domain);
\path (integral_domain) edge (domain);
\path (integral_domain) edge (nonzero_comm_ring);
\path (local_ring) edge (nonzero_comm_ring);
\path (topological_ring) edge[bend right=5] (topological_add_group);
\path (uniform_add_group) edge[bend left=25] (topological_add_group);
\path (principal_ideal_domain) edge (integral_domain);
\path (principal_ideal_domain) edge (is_noetherian_ring);
\path (metric_space) edge (has_dist);
\path (metric_space) edge (emetric_space);
\path (topological_semiring) edge (topological_monoid);

\node (topological_add_monoid) at (-3.9, -6) {\lstinline{topological_add_monoid}};
\node (has_edist) at (-7.5, -2.6) {\lstinline{has_edist}};
\node (first_countable_topology) at (-7.3, -3.2) {\lstinline{first_countable_topology}};
\node (separated) at (-4.5, -3) {\lstinline{separated}};
\node (no_zero_divisors) at (1.9, -3) {\lstinline{no_zero_divisors}};
\node (nonzero_comm_semiring) at (5, -3) {\lstinline{nonzero_comm_semiring}};
\node (comm_ring) at (8, -3) {\lstinline{comm_ring}};
\path (nonzero_comm_ring) edge (comm_ring);
\path (nonzero_comm_ring) edge (nonzero_comm_semiring);
\path (domain) edge (no_zero_divisors);
\path (topological_semiring) edge (topological_add_monoid);
\path (topological_add_group) edge (topological_add_monoid);
\path (emetric_space) edge[bend right=10] (has_edist);
\path (emetric_space) edge (separated);
\path (emetric_space) edge (first_countable_topology);

\node (sequential_space) at (-7.3, -7) {\lstinline{sequential_space}};
\node (uniform_space) at (-4, -4) {\lstinline{uniform_space}};
\node (regular_space) at (-6.2, -4) {\lstinline{regular_space}};
\node (zero_ne_one_class) at (2.5, -4) {\lstinline{zero_ne_one_class}};
\node (ring) at (5, -4) {\lstinline{ring}};
\node (comm_semiring) at (7, -4) {\lstinline{comm_semiring}};
\path (comm_ring) edge (comm_semiring);
\path (nonzero_comm_semiring) edge (comm_semiring);
\path (comm_ring) edge (ring);
\path (domain) edge[bend left=20] (ring);
\path (domain) edge[bend left=10] (zero_ne_one_class);
\path (nonzero_comm_semiring) edge (zero_ne_one_class);
\path (topological_ring) edge (ring);
\path (is_noetherian_ring) edge[bend left=3] (ring);
\path (first_countable_topology) edge (sequential_space);
\path (separated) edge (regular_space);
\path (separated) edge (uniform_space);
\path (uniform_add_group) edge[bend left=10]  (uniform_space);

\node (t2_space) at (-6.5, -5) {\lstinline{t2_space}};
\node (add_comm_group) at (2, -5) {\lstinline{add_comm_group}};
\node (semiring) at (4, -5) {\lstinline{semiring}};
\node (has_dvd) at (6, -5) {\lstinline{has_dvd}};
\node (comm_monoid) at (8, -5) {\lstinline{comm_monoid}};
\path (comm_semiring) edge (has_dvd);
\path (ring) edge (semiring);
\path (comm_semiring) edge (semiring);
\path (comm_semiring) edge (comm_monoid);
\path (ring) edge (add_comm_group);
\path (normed_group) edge[bend left=5] (add_comm_group);
\path (topological_semiring) edge (semiring);
\path (regular_space) edge (t2_space);

\node (add_group) at (-0.5, -6) {\lstinline{add_group}};
\node (add_comm_monoid) at (1.5, -6) {\lstinline{add_comm_monoid}};
\node (mul_zero_class) at (3.7, -6) {\lstinline{mul_zero_class}};
\node (distrib) at (5.5, -6) {\lstinline{distrib}};
\node (comm_semigroup) at (8.5, -6) {\lstinline{comm_semigroup}};
\node (monoid) at (6.8, -6) {\lstinline{monoid}};
\node (t1_space) at (-6.5, -6) {\lstinline{t1_space}};
\path (semiring) edge (distrib);
\path (semiring) edge (mul_zero_class);
\path (semiring) edge (add_comm_monoid);
\path (add_comm_group) edge (add_comm_monoid);
\path (semiring) edge (monoid);
\path (comm_monoid) edge (monoid);
\path (topological_monoid) edge[bend left=4] (monoid);
\path (comm_monoid) edge (comm_semigroup);
\path (add_comm_group) edge (add_group);
\path (topological_add_group) edge (add_group);
\path (t2_space) edge (t1_space);

\node (add_comm_semigroup) at (2, -7) {\lstinline{add_comm_semigroup}};
\node (add_monoid) at (-0.3, -7) {\lstinline{add_monoid}};
\node (add_left_cancel_semigroup) at (5.3, -7)
{\begin{tabular}{c}\lstinline{add_left_cancel_semigroup} \&\\\lstinline{add_right_cancel_semigroup}\end{tabular}};
\node (semigroup) at (8.5, -7) {\lstinline{semigroup}};
\node (has_neg) at (-2, -7) {\lstinline{has_neg}};
\node (has_sub) at (-1.5, -7.5) {\lstinline{has_sub}};
\node (has_one) at (7.7, -7.5) {\lstinline{has_one}};
\node (t0_space) at (-5.3, -7) {\lstinline{t0_space}};

\path (add_group) edge (has_sub);
\path (add_group) edge (has_neg);
\path (comm_semigroup) edge (semigroup);
\path (monoid) edge (semigroup);
\path (add_comm_monoid) edge (add_monoid);
\path (add_group) edge (add_monoid);
\path (add_comm_monoid) edge (add_comm_semigroup);
\path (add_group) edge (add_left_cancel_semigroup);
\path (monoid) edge[bend left=10] (has_one);
\path (topological_add_monoid) edge[bend left=5] (add_monoid);
\path (t1_space) edge (t0_space);

\node (add_semigroup) at (2, -8) {\lstinline{add_semigroup}};
\node (has_zero) at (-0.3, -8) {\lstinline{has_zero}};
\node (has_mul) at (8.5, -8) {\lstinline{has_mul}};
\node (topological_space) at (-4, -8) {\lstinline{topological_space}};

\path (add_monoid) edge (add_semigroup);
\path (add_monoid) edge (has_zero);
\path (semigroup) edge (has_mul);
\path (add_comm_semigroup) edge (add_semigroup);
\path (add_left_cancel_semigroup) edge (add_semigroup);
\path (t0_space) edge (topological_space);
\path (topological_monoid) edge (topological_space);
\path (topological_add_monoid) edge[bend right=15] (topological_space);
\path (sequential_space) edge (topological_space);
\path (uniform_space) edge[bend left=10] (topological_space);

\node (has_add) at (2, -8.7) {\lstinline{has_add}};
\node (measurable_space) at (-4, -8.7) {\lstinline{measurable_space}};
\path (add_semigroup) edge (has_add);
\path (topological_space) edge (measurable_space);

 \end{tikzpicture}
 }

 \caption{The structure hierarchy underlying \lstinline{normed_field}.
 An edge from \lstinline{S} to \lstinline{T} indicates that an instance of \lstinline{T} can be derived from an instance of \lstinline{S}.
 Some arrows to the leaves are omitted for readability (e.g. from \lstinline{zero_ne_one_class} to \lstinline{has_one}).}
 \label{figure:structure-example}
\end{figure*}

A fragment of the structure hierarchy in Figure~\ref{figure:structure-example} shows the classes that can be derived from a \lstinline{normed_field} instance, which appears near the top of the hierarchy.
The bottom of the graph is populated by notation classes such as \lstinline{has_add}, \lstinline{has_le}, \lstinline{has_one}, etc.
The fragment omits structures such as \lstinline{partial_order} and \lstinline{lattice}, as well as ordered structures like \lstinline{ordered_group} and \lstinline{ordered_ring}.

Most of the classes in the hierarchy are unary: they have the form $C\;\alpha$ where $\alpha$ is the carrier type.
All of the classes in the fragment above are unary, but there are some binary classes, such as \lstinline{module R M} that provides an $R$-module structure on $M$.
The fragment constitutes less than a third of the current structure hierarchy, which contains 201 unary classes and 266 instances of the form $C \; \alpha \to D \; \alpha$, which only change the class associated to a type.

Except for notation classes, all the classes in \mathlib have some independent interest.
Uniform spaces are defined because they unify theorems of topological groups and metric spaces; extended metric spaces exist because they appear in constructions such as the unbounded $\ell^p$ spaces.
This does not hold for some \corelib\ classes, e.g.\ \lstinline{zero_ne_one_class}, which exist only to be mixins for other classes.

Diamonds are common---the structure hierarchy is far from being a tree.
Some of these diamonds arise naturally from the mathematics, some arise when a class is defined as a least upper bound of two other classes (such as \lstinline{comm_monoid}), and others arise as a result of encoding a canonical projection as a parent class.

In some cases, we have a situation in which one structure canonically implies another: for example, a metric space is not generally considered to \emph{contain} a topology as a sub-component, but the metric uniquely defines a particular topology of interest.
We opt to have \lstinline{metric_space} extend \lstinline{topological_space}, with an additional constraint asserting that the topology agrees with the induced topology of the metric.
This is done because Lean depends on definitional equality of projections and this approach makes more squares commute definitionally.
Using this approach, the induced topology on a product metric becomes definitionally equal to the product topology on the induced metrics.

\subsubsection{Bundled Type Classes}

When creating a type class, one important design decision is which parts of the definition to put as parameters to the type class and which parts to store within the element itself, accessible via a projection.
We refer to a type class as \emph{unbundled} if it has many parameters and \emph{bundled} if the parameters are moved to projections.
For example, given the definitions
\begin{enumerate}
\item $\mathsf{Group}=\{(X,\circ)\mid (X,\circ)\mbox{ is a group}\}$
\item $\mathsf{group}\;X=\{\circ\mid (X,\circ)\mbox{ is a group}\}$
\item $\mathsf{is\_group}\;X\;\circ\;\leftrightarrow\; (X,\circ)\mbox{ is a group}$
\end{enumerate}
we would call $\mathsf{Group}$ a bundled definition, $\mathsf{is\_group}$ an unbundled definition, and $\mathsf{group}$ a semi-bundled definition.
All of these say essentially the same thing from a mathematical point of view, but the choice of which to use has a significant impact on formalization.

We primarily use semi-bundled definitions for type classes in \mathlib, with all operations being bundled except for the carrier types.
For use with type classes, fully bundling (as with $\mathsf{Group}$) is not an option: a type class problem should have (essentially) at most one solution, and only the parameters affect the type class search.
When the type is exposed but not the operation, as with $\mathsf{group}\;X$, we can register a canonical group associated to the type $X$ if there is one.
For instance, $\mathsf{add\_group}\;\mathbb{Z}$ will find the canonical additive group $(\mathbb{Z},+)$ of addition on integers.
By contrast,
fully bundled definitions work best when using canonical structures,
as in the \mathcomp library \cite{Mahb17}.

The drawback of further unbundling,
as in $\mathsf{is\_group}\;X\;\circ$,
is that it makes matching problems difficult.
For example, a theorem saying that $f\;x\;y=f\;y\;x$
under the condition $\mathsf{is\_comm\_group}\;X\;f$,
with $x,y,f$ all variables,
leads to a higher-order matching problem:
$f$ could be substituted for a lambda term.
This kind of theorem will always be applicable,
since essentially any term can be written in the form $f\;x\;y$ for some choice of $f$.
Especially when the simplification tactic \lstinline{simp} is invoked,
this can cause a significant performance problem,
with many false positives and failed type class searches.
With the partially unbundled approach this statement instead has the form $\mathsf{add}\;X\;m\;x\;y=\mathsf{add}\;X\;m\;y\;x$ where $m:\mathsf{comm\_group}\;X$.
We need only look for terms which have a literal appearance of the constant $\mathsf{add}$ in them rather than considering all terms and excluding them after the fact by a failed type class search.

\subsubsection{Bundled Morphisms}

A similar distinction arises when talking about morphisms between structures.
In most cases, a morphism will be a function with an additional property, for example a continuous function or a group homomorphism.
In these cases we have two options: to bundle the function with the property, producing a type $\mathrm{Hom}(A,B)$ of structure-respecting functions from $A$ to $B$, or to have a function $f:A\to B$ and a type class $\mathsf{is\_hom}\;f$ asserting that $f$ respects the structure of $A$ and $B$.

Both options are workable, and \mathlib contains traces of both approaches.
However, we have found the bundled approach to behave better for a number of reasons:

\begin{itemize}
\item It is important for compositions of homs to be a hom,
but problems of the form $\mathsf{is\_hom}\;(f\circ g)$ can be difficult for type class inference.
With bundled homs, one instead defines a custom composition operator $\mathrm{Hom}(A,B)\times \mathrm{Hom}(B,C)\to \mathrm{Hom}(A,C)$ that is used explicitly, obviating the need to search for $\mathsf{is\_hom}\;(f\circ g)$.
\item The type $\mathrm{Hom}(A,B)$ itself often has interesting structure.
For example, the type \lstinline{M →ₗ[R] N} of $R$-linear maps from $M$ to $N$ is an $R$-module where the zero element is the constant zero map, and addition and scalar multiplication work pointwise (Section~\ref{subsection:linalg:module}).
Bundling the type makes it easier to reason about this structure.
\end{itemize}

\label{subsection:typeclasses:bundling}

\subsection{The \texttt{decidable} Type Class}
\label{subsection:typeclasses:decidability}

The class \lstinline{decidable p} is defined essentially as the disjunctive type \lstinline{p ⊕ ¬ p}: it is a constructive, \lstinline{Type}-valued witness to the law of excluded middle at proposition \lstinline{p}.
Although \mathlib primarily deals in classical mathematics, it is  useful to track decidability because it can be used in algorithms and for small scale computation in the executable fragment.

The notation \lstinline{if p then t else e} is syntactic sugar for \lstinline{ite p t e}, where:
\begin{lstlisting}
ite : ∀ (p : Prop) [d : decidable p],
  ∀ {α : Sort u}, α → α → α
\end{lstlisting}
This function is defined by case analysis on the witness \lstinline{d}.
In Lean syntax, parentheses~\lstinline{()} denote explicit arguments,
curly brackets~\lstinline|{}| implicit arguments,
and square brackets~\lstinline{[]} arguments to be inferred by type class resolution.
Notice that \lstinline{p} is an explicit argument, but \lstinline{p}, being a proposition, is erased during execution, with the real condition in \lstinline{d}.
Morally, we can think of \lstinline{d} as a boolean value, but it carries evidence with it connecting it to the truth or falsity of \lstinline{p}.

The \mathcomp library~\cite{Mahb17} relies heavily on the technique of \emph{small-scale reflection} (from which the tactic language SSReflect gets its name), where many predicates and propositions are defined to live in type \lstinline{bool}. The computational behavior of the logic can be used to discharge certain proof obligations by reflection.
This means that most properties have two versions, one that is a \lstinline{Prop} and one which is a \lstinline{bool}, with an additional predicate \lstinline{reflect b p} asserting that the boolean value \lstinline{b} is true iff \lstinline{p} holds.

A proof of \lstinline{decidable p} is equivalent to \lstinline{Σ b,reflect b p} in this sense, but because it is a type class the management of this side condition is almost entirely automatic.
This means we can treat the proposition \lstinline{p} as primary, as in the definition of \lstinline{ite}, and need not define two versions of every proposition.
We can still prove true decidable propositions by computational reflection via \lstinline{dec_trivial}:
\begin{lstlisting}
def as_true (c : Prop) [decidable c] : Prop :=
if c then true else false
\end{lstlisting}
\begin{lstlisting}
def of_as_true {c : Prop} [decidable c] :
  as_true c → c := ...
\end{lstlisting}
\begin{lstlisting}
notation `dec_trivial` := of_as_true trivial
\end{lstlisting}

The considerations in this section apply equally to proving, programming, and metaprogramming in Lean.
All propositions can be shown to be (noncomputably) decidable with the \lstinline{choice} axiom;
this instance is used locally with low priority in \mathlib
to use declarations with decidability hypotheses in classical mathematics.

\subsection{Problems with Type Classes}
\label{subsection:typeclasses:bad}

Given all the aforementioned uses of type classes, 
and the fact that most of instances are available in most of the higher-level files, 
the growth of \mathlib has tested the scalability of the type class inference system.
A file that imports all of \mathlib has access to 4256 instances among 386 classes.
While it has held up remarkably well, aspects of the inference algorithm have lead to limitations in some areas and performance walls in others.

First, Lean performs a backtracking search on every type class problem.
Thus, an instance such as \lstinline{C α ← C α} (that is, ``to obtain \lstinline{C α} it suffices to prove \lstinline{C α}'') will cause the inference routine to enter a loop, blocking resolution even if a successful path exists elsewhere.
As a result, \mathlib requires the complete instance graph to be acyclic.
This is a global problem but generally single instances can be identified as the culprit when an instance loop is discovered.

Second, and relatedly, because Lean cannot tell when it is retreading the same paths, even if the graph of instances has no cycles, it will traverse all paths through the graph, which is exponential time in the worst case.
(On successes it is often significantly less if it gets lucky, but failures will have to search the whole graph, and pathological examples can be constructed in which the search is successful but still exponential time.) Instance searches can take hundreds of thousands of steps through our thousands of instances.

Third, instance searches are performed exclusively backward, reasoning from the goal back through instances.
This means that if for example \lstinline{G : group α} is in the context,
a search for \lstinline{has_mul α} may end up exploring upward through the entire structure hierarchy, looking for \lstinline{monoid α}, then \lstinline{ring α}, then \lstinline{field α}, then \lstinline{normed_field α}, prompting a search for \lstinline{has_norm α} and so on, before eventually failing and attempting \lstinline{group α} instead.
This search will get larger as our structure hierarchy grows.

While the first two problems seem dire, a combination of good caching and a very efficient C++ implementation have helped to keep costs manageable even at \mathlib's scale.
Moreover, efficient graph traversal algorithms are known, and Lean 4 is expected to make improvements in this area.

The third problem could potentially be fixed by using a combination of forward and backward reasoning.
If we reason forward from \lstinline{group α} to \lstinline{monoid α}, \lstinline{semigroup α}, \lstinline{has_mul α}, then work backward from the goal \lstinline{has_mul α}, we find the solution quickly without exploring the entire structure hierarchy.
Working forward from instances such as \lstinline{ring ℤ} entails generating additional instances for \lstinline{monoid ℤ}, \lstinline{add_group ℤ}, and so on, which is currently not required but is often done to speed up instance searches.

\section{Linear Algebra}
\label{section:linalg}

The development of linear algebra in \mathlib showcases many of the design principles explained so far.
To make earlier discussions more concrete,
we describe in detail part of the linear algebra development.
This description does not encompass all of the linear algebra in \mathlib:
the library also develops the theories of dual vector spaces, bilinear and sesquilinear forms, direct sums, and tensor products.
This theory contains structures and theorems that are mathematically nontrivial
and are used to prove noteworthy results~(Section~\ref{subsection:community:oneoff}).
Some of the structures are close to the top of the type class hierarchy,
and they illustrate the use of bundled and semi-bundled type classes and quotient types.

Our formalization is certainly not unique:
we have been heavily inspired by existing linear algebra developments,
particularly in Isabelle/HOL
\cite{Lee14, Diva13, Aran14}
and Coq \cite{Gont11}.

\subsection{Modules}
\label{subsection:linalg:module}

The fundamental structures of linear algebra are \emph{modules}.
Given a ring $R$,
an $R$-module consists of
an abe\-lian group $M$ and
a distributive, associative scalar multiplication operator $\smul : R \times M \to M$.
In \mathlib,
the type class \lstinline{module R M} defines an \lstinline{R}-module structure on the group \lstinline{M}.

\begin{lstlisting}
class mul_action (α : Type u) (β : Type v)
  [monoid α] extends has_scalar α β :=
(one_smul : ∀ b : β, (1 : α) • b = b)
(mul_smul : ∀ (x y : α) (b : β),
            (x * y) • b = x • y • b)
\end{lstlisting}
\begin{lstlisting}
class distrib_mul_action (α : Type u)
  (β : Type v) [monoid α] [add_monoid β]
  extends mul_action α β :=
(smul_add : ∀ (r : α) (x y : β),
            r • (x + y) = r • x + r • y)
(smul_zero : ∀ (r : α), r • (0 : β) = 0)
\end{lstlisting}
\begin{lstlisting}
class semimodule (R : Type u) (M : Type v)
  [semiring R] [add_comm_monoid M]
  extends distrib_mul_action R M :=
(add_smul : ∀ (r s : R) (x : M),
            (r + s) • x = r • x + s • x)
(zero_smul : ∀x : M, (0 : R) • x = 0)
\end{lstlisting}
\begin{lstlisting}
class module (R : Type u) (M : Type v)
  [ring R] [add_comm_group M]
  extends semimodule R M
\end{lstlisting}
We provide an alternative constructor for \lstinline{module R M}
that does not require the fields \lstinline{smul_zero} and \lstinline{zero_smul},
since they follow from other properties.

When $R$ is replaced with a field $K$,
this structure is called a $K$-vector space.
All results about modules also apply to vector spaces.
In \mathlib, we favor working with discrete fields,
which fix $1/0=0$.

\begin{lstlisting}
class vector_space (K : Type u) (V : Type v)
  [discrete_field K] [add_comm_group V]
  extends module K V
\end{lstlisting}

We consider $R$-modules for the rest of this section.
For the sake of display, 
we will omit the following parameters to all declarations:
\begin{lstlisting}
(R : Type u) (M : Type v) [ring R]
[add_comm_group M] [module R M]
\end{lstlisting}
We will also sometimes use ``\lstinline{...}'' to elide proof terms.

When $(M, \smul)$ is an $R$-module,
a subgroup of $M$ closed under~$\smul$ is called a \emph{submodule}.
These are also defined as bundled structures in \mathlib.
Ordered by set inclusion, the submodules of a module form a complete lattice.

\begin{lstlisting}
structure submodule :=
(carrier : set M)
(zero : (0:M) ∈ carrier)
(add  : ∀ {x y}, x ∈ carrier →
        y ∈ carrier → x + y ∈ carrier)
(smul : ∀ (c:R) {x},
        x ∈ carrier → c • x ∈ carrier)
\end{lstlisting}
The lattice structure on submodules allows us to use
the notation \lstinline{⊥} for the trivial submodule ${0}$
and \lstinline{⊤} for the universal submodule $M$.
We also use the lattice structure to define the \emph{span} of a set,
the space of all linear combinations of its elements:

\begin{lstlisting}
def span (s : set M) : submodule R M :=
Inf {p | s ⊆ p}
\end{lstlisting}

The \emph{quotient} $M/N$ of an $R$-module $M$ by a submodule $N \subseteq M$
equates $m_1,m_2 \in M$ if $m_1 - m_2 \in N$.
The quotient is itself an $R$-module,
and is defined in \mathlib as a quotient type.

\begin{lstlisting}
def quotient_rel (N : submodule R M) :
  setoid M := ⟨λ x y, x - y ∈ N, ...⟩
\end{lstlisting}
\begin{lstlisting}
def submodule.quotient (N : submodule R M) :
  Type v := quotient (quotient_rel N)
\end{lstlisting}
\begin{lstlisting}
instance (N : submodule R M) :
  module R (quotient N) := ...
\end{lstlisting}

A function between two $R$-modules that preserves addition and scalar multiplication
is called a \emph{linear map}.
If it is invertible, it is called a \emph{linear equivalence}.
We work with linear maps as bundled structures in \mathlib,
and coercions typically allow us to treat these structures as functions.
We use the notation \lstinline{M →ₗ[R] N} to stand for \lstinline{linear_map R M N}
and \lstinline{M ≃ₗ[R] N} for \lstinline{linear_equiv R M N}.

\begin{lstlisting}
structure linear_map (N : Type w)
  [add_comm_group N] [module R N] :=
(to_fun : M → N)
(add  : ∀ x y,
        to_fun (x + y) = to_fun x + to_fun y)
(smul : ∀ (c : R) x,
        to_fun (c • x) = c • to_fun x)
\end{lstlisting}
The types \lstinline{M →ₗ[R] N} and \lstinline{M ≃ₗ[R] N} are themselves both $R$-modules.

If $f : M \to N$ is linear,
the image of a submodule $M' \subseteq M$ under $f$ is a submodule of $N$,
and the preimage of a submodule $N' \subseteq M$ under $f$ is a submodule of $M$.
These are defined respectively as \lstinline{map} and \lstinline{comap}.
In particular, the \emph{kernel} of $f$---defined to be the set
$\left\{ x \in M \mid f(x)=0 \right\}$---is a submodule, as is the range.

\begin{lstlisting}
def ker (f : M →ₗ[R] N) : submodule R M :=
comap f ⊥
\end{lstlisting}
\begin{lstlisting}
def range (f : M →ₗ[R] N) : submodule R N :=
map f ⊤
\end{lstlisting}

These definitions and their surrounding proofs
are enough to prove the first and second \emph{isomorphism laws} for modules:

\begin{lstlisting}
def quot_ker_equiv_range :
  f.ker.quotient ≃ₗ[R] f.range := ...
\end{lstlisting}
\begin{lstlisting}
def sup_quotient_equiv_quotient_inf
  (p p' : submodule R M) :
  (comap p.subtype (p ⊓ p')).quotient ≃ₗ[R]
    (comap (p ⊔ p').subtype p').quotient := ...
\end{lstlisting}

\subsection{Bases}
\label{subsection:linalg:bases}

A set of elements of an $R$-module $\left\{ v_i \right\} _{i \in I}$
is said to be \emph{linearly dependent} if
there is a finite subset $I' \subseteq I$
and a family of scalars $\left\{ c_i \right\} _{i \in I'}$, not all zero, such that
$\sum_{i \in I'}c_i \smul v_i = 0$.
We are more often interested in the negation of this notion.
In \mathlib, we define the linear independence of a family of vectors
indexed by a base type:
\begin{lstlisting}
def linear_independent
  {ι : Type u} (R : Type v) {M : Type w}
  [ring R] [add_comm_group M] [module R M]
  (v : ι → M) : Prop :=
(finsupp.total ι M R v).ker = ⊥
\end{lstlisting}
The function \lstinline{finsupp.total ι M R v} \lstinline{:} \lstinline{(ι →₀ R) →ₗ[R] M}
sends a finitely supported function \lstinline{c : ι → R}
to the sum over \lstinline{ι} of \lstinline{c i • v i}.

Through the rest of this subsection
we fix parameters as given to \lstinline{linear_independent}.
A linearly independent family of vectors is a \emph{basis} for a module
if it spans the entire module.

\begin{lstlisting}
def is_basis (v : ι → M) : Prop :=
linear_independent R v ∧ span R (range v) = ⊤
\end{lstlisting}

Given such a \lstinline{v} and a proof \lstinline{hv} \lstinline{:} \lstinline{is_basis R v},
we define the linear map that decomposes a vector into a weighted sum of vectors in \lstinline{v}:

\begin{lstlisting}
def is_basis.repr : M →ₗ (ι →₀ R) :=
hv.1.repr.comp
  (linear_map.id.cod_restrict _ hv.mem_span)
\end{lstlisting}

A classic result in linear algebra shows that every vector space has a basis.
(This result uses Zorn's lemma and does not hold constructively.)

\begin{lstlisting}
lemma exists_is_basis [discrete_field K]
  [add_comm_group V] [vector_space K V] :
  ∃b : set V, is_basis K (λ i : b, i.val) := ...
\end{lstlisting}

\subsection{Dimension}
\label{subsection:linalg:dimension}

Any two bases for a vector space $V$ have equal cardinality.
This cardinality defines the \emph{dimension} of $V$.
Vector spaces are not necessarily finite dimensional,
so we define the dimension function to take values in the type \lstinline{cardinal}.
The theory of cardinal numbers in \mathlib is located in the \lstinline{set_theory} subfolder.
To avoid inconsistency, the type \lstinline{cardinal} must be parametrized by a universe level.

\begin{lstlisting}
def vector_space.dim (K : Type u) (V : Type v)
  [discrete_field K] [add_comm_group V]
  [vector_space K V] : cardinal.{v} :=
cardinal.min (nonempty_subtype.2
   			   (@exists_is_basis K V _ _ _))
 			      (λ b, cardinal.mk b.1)
\end{lstlisting}

Because the relevant cardinals may live in different universes,
it takes some care to state the dimension theorem.
\begin{lstlisting}
theorem mk_eq_mk_of_basis
  {v : ι → V} {v' : ι' → V}
  (hv : is_basis K v) (hv' : is_basis K v') :
  cardinal.lift.{w w'} (cardinal.mk ι) =
    cardinal.lift.{w' w} (cardinal.mk ι') := ...
\end{lstlisting}

Many dimension computations are proved in \mathlib,
including results about the dimensions of quotients
and the rank-nullity theorem.

\begin{lstlisting}
theorem dim_quotient (V' : submodule K V) :
  dim K p.quotient + dim K V' = dim K V := ...
\end{lstlisting}
\begin{lstlisting}
theorem dim_range_add_dim_ker (f : V →ₗ[K] E) :
  dim K f.range + dim K f.ker = dim K V := ...
\end{lstlisting}
A number of these computations specialize to finite-dim\-ensional vector spaces.
For instance, the functions $A \to k$ form a $k$-vector space
with dimension $|A|$ when $A$ is finite.

\begin{lstlisting}
lemma dim_fun [fintype η] :
  dim K (η → K) = fintype.card η := ...
\end{lstlisting}
\begin{lstlisting}
lemma dim_fin_fun (n : ℕ) :
  dim K (fin n → K) = n := ...
\end{lstlisting}
The vector space \lstinline{fin n → K} is a common way to represent \lstinline{n}-tuples of elements of \lstinline{K}.

\subsection{Matrices}
\label{subsection:linalg:matrices}

It is convenient to consider matrices as functions $m \to n \to R$, where $m$ and $n$ are finite sets of indices.
We define this type in the \lstinline{data} directory of \mathlib,
along with the expected operations and algebraic instances.
For a ring $R$, $R$-valued $m \times n$ matrices form an $R$-module.
\begin{lstlisting}
def matrix (m n : Type u) (R : Type v)
  [fintype m] [fintype n] : Type (max u v) :=
m → n → R
\end{lstlisting}
\begin{lstlisting}
instance [ring R] : module R (matrix m n R) :=
...
\end{lstlisting}
There is a direct correspondence between $R$-valued matrices and linear maps between $R$-modules.
To begin with, every such $m \times n$ matrix gives rise to a linear map from $R^n$ to $R^m$.
In fact, this evaluation function is itself a linear map.
\begin{lstlisting}
def eval : (matrix m n R) →ₗ[R]
  ((n → R) →ₗ[R] (m → R)) :=
linear_map.mk₂ R mul_vec ...
\end{lstlisting}

Similarly, a linear map exists in the opposite direction.
\begin{lstlisting}
def to_matrix : ((n → R) →ₗ[R] (m → R))
  →ₗ[R] matrix m n R :=
linear_map.mk
  (λ f i j, f (λ n, ite (j = n) 1 0) i) ...
\end{lstlisting}

These linear maps are inverses of each other,
and thus form an equivalence between
the space of linear maps $R^n \to R^m$
and the space of $R$-valued $m \times n$ matrices.
It follows quickly that,
given finite bases for two $R$-modules $M_1$ and $M_2$ indexed by $\iota$ and $\kappa$,
the space of linear maps $M_1 \to M_2$ is linearly equivalent to the space of $R$-valued $\iota \times \kappa$ matrices.
\begin{lstlisting}
def lin_equiv_matrix {ι : Type i} {κ : Type k}
  [fintype ι] [decidable_eq ι]
  [fintype κ] [decidable_eq κ]
  {v₁ : ι → M₁} (hv₁ : is_basis R v₁)
  {v₂ : κ → M₂} (hv₂ : is_basis R v₂) :
  (M₁ →ₗ[R] M₂) ≃ₗ[R] matrix κ ι R := ...
\end{lstlisting}

\subsection{Summary}
\label{subsection:linalg:summary}

While \mathlib's algebraic content is significant, 
it is by no means the majority of the library, 
and we do not mean to imply this by focusing on linear algebra as an example.
The preceding sections intend to give the reader an idea of the flavor of \mathlib formalizations more generally. 
The use of type classes to manage algebraic and order structure is representative of the whole library.

\section{Metaprogramming}
\label{section:metaprogramming}

Due to Lean's powerful metaprogramming framework~\cite{EURAM17}, many features
that might otherwise require changes to the prover or extension
through plugins are implemented in \mathlib itself. These
include general purpose decision procedures, utilities, debugging
tools, and special purpose automation.
All of the following tactics are implemented in Lean as metaprograms,
with the exceptions of the core simplification routine,
E-matching~\cite{Mour07},
and congruence closure~\cite{Sels16},
which are exposed in the metalanguage as atomic procedures.

\subsection{Simplification}
\label{subsection:metaprogramming:simp}

The tactic \lstinline{simp} is the primary tool
in the Lean automation arsenal.
Similar to the simplifier in Isabelle,
it performs non-definitional directional rewriting with equality and biconditional lemmas.
Lean's \emph{attribute} mechanism allows the user
to annotate definitions and theorems with extra information,
which can be accessed by metaprograms;
the default set of rewrite rules for \lstinline{simp} is extended
by adding the \lstinline{simp} attribute to declarations.
A variant, \lstinline{dsimp}, performs only definitional reductions.

The simplifier matches rewrite lemmas up to syntactic, not definitional, equality.
For this reason, lemmas in \mathlib are typically stated in \lstinline{simp}-normal form.
When there are multiple equivalent ways to express a term,
one is preferred, and others are simplified to the preferred form;
subsequent lemmas need only be stated for this single case.
An example of this design pattern can be seen in the development of the $p$-adic numbers \lstinline{ℚ_[p]},
where the generic norm notation \lstinline{∥x∥}
from the \lstinline{normed_space} type class
is preferred to the definitionally equal \lstinline{padic_norm p x}.
We use \lstinline{simp} lemmas to transform the latter to the former,
and develop the library for the $p$-adic norm based on the generic notation.
\begin{lstlisting}
@[simp] lemma is_norm (q : ℚ_[p]) :
  (padic_norm_e q) = ∥q∥ := rfl
\end{lstlisting}
\begin{lstlisting}
@[simp] lemma mul (q r : ℚ_[p]) :
  ∥q * r∥ = ∥q∥ * ∥r∥ := ...
\end{lstlisting}

\subsection{Tactics and Automation}
\label{subsection:metaprogramming:tactics}

Besides \lstinline{simp}, we briefly describe notable examples of
automation in \mathlib, which we roughly classify according to whether
they are ``big'' (general-purpose, powerful, and meant to discharge
certain classes of goals on their own) or ``small'' (specialized and
providing finer control over the proof state).
Alongside tactics, we include user commands,
which manipulate the prover environment outside of any particular proof state,
and attributes,
which can trigger metaprograms when applied to declarations.


\paragraph{Big Automation} For general-purpose proof search,
the most powerful tools in \mathlib are \lstinline{finish} and \lstinline{tidy}.
The former combines a tableau prover (complete for  propositional logic)
with simplification, E-matching, and congruence closure.
The latter implements heuristics
and repeatedly attempts to apply a preselected list of tactics (recursing into subgoals)
until none succeed.
In the \texttt{category theory} library, \lstinline{tidy} is used extensively
to check functoriality and naturality.
The constructors for many structures call \lstinline{tidy} by default,
so that proofs of these ``obvious'' properties need not even be mentioned.

The tactics \lstinline{ring} and \lstinline{abel}
normalize expressions in commutative (semi)rings and abelian groups.
They follow the approach described by Gr\'egoire and Mahboubi~\cite{Greg05},
but produce proof terms tracing each normalization step
instead of verifying by reflection.

Two tactics are used for solving linear inequalities.
For linear ordered commutative semirings,
\lstinline{linarith} implements an algorithm based on Fourier-Motzkin variable elimination~\cite{Will86};
it is complete for dense linear orders.
On $\mathbb{N}$ and $\mathbb{Z}$,
\lstinline{omega} partially implements the omega decision procedure
for Presburger arithmetic~\cite{Baek19, Pugh91}. 


\paragraph{Small Automation} While Lean's metaprogramming engine is powerful enough for large tactics,
it shines in the context of special-purpose tools,
which are often simple to create and deploy.
We highlight here a few of the dozens of such tactics and commands implemented in \mathlib.

The \lstinline{norm_cast} tactic manipulates
type coercions \cite{Made19}. Because Lean has no subtyping, a coercion
(written with a prefix \lstinline{↑})
is required, for instance, when using a
natural number in place of an integer. The presence of these coercions in terms can
hinder simplification and unification:
it can be tedious to reduce the integer inequality \lstinline{↑m + ↑n > 5}
to the natural number inequality \lstinline{m + n > 5}.
Using attributes to track lemmas that pass coercions through operations and relations,
\lstinline{norm_cast} performs such simplification automatically.

Arithmetic expressions involving only literals
are evaluated efficiently by \lstinline{norm_num}.
Over the natural numbers,
goals such as \lstinline{1 + 2 < 4}
can be proved by kernel computation.
However, this is inefficient for large expressions,
and impossible on noncomputable types such as \lstinline{ℝ}
and on arbitrary linear ordered semirings.
As an alternative,
\lstinline{norm_num} uses the syntactic binary representation of numerals
and lemmas about the relevant algebraic structures
to simplify the arithmetic expressions as far as possible.


For a type class \lstinline{C},
the tactic \lstinline{pi_instance} generalizes instances of the form
\lstinline{C α} to \lstinline{C (β → α)}.
With this, we obtain much of the algebraic structure of function spaces for free.

The \lstinline{reassoc} attribute improves the ability of \lstinline{simp} to
reason about compositions of arrows in a category modulo associativity.
By default, \lstinline{simp} normalizes
such expressions by associating composition to the right. As a
consequence, one often encounters series of arrow compositions that
are bracketed as \lstinline{a ≫ (b ≫ c)}.
A lemma \lstinline{foo}
that rewrites \lstinline{a ≫ b} to \lstinline{d}
would conflict with this \lstinline{simp} normal form.
Applying the attribute \lstinline{reassoc} to \lstinline{foo}
produces a companion lemma
of the shape \lstinline{∀ X, a ≫ (b ≫ X) = d ≫ X}
that can be used by the simplifier.

%


\paragraph{Maintenance and Fine-Tuning} Another group of meta\-programs is used for library development and maintenance.

Some search-based tactics can report traces of a successful search.
Variants of \lstinline{tidy} and \lstinline{simp}
list the tactic script and used lemmas, respectively,
in a format that is faster and more robust than the original tactic call.
In editors that support Lean's \emph{hole commands},
calls to these tactics can be literally self-replacing.

The command \lstinline{#lint} is a linting tool
that performs various tests on declarations in a file or environment.
Among other things, it checks for unused arguments, malformed names, and
whether a file meets \mathlib's documentation requirements.
It is easily extended with custom tests.

Lean's namespacing and sectioning mechanisms
allow the use of abbreviated identifiers, special notation, and instances
locally in files,
as well as to automatically insert parameters to declarations within a section.
The \lstinline{#where} command prints information about the currently open namespaces and parameters;
the \lstinline{localized} command associates local notation and instances with a namespace, making it easy to set up a particular local environment.

\subsection{Lessons}

The various tools listed above
are provided as a part of \mathlib,
without need for extensions or plugins to the Lean system.
(One exception, namely \lstinline{simp}, is built into Lean.)

Collaboration between mathematically-oriented formalizers
and more experienced programmers
has driven much of the tactic development in \mathlib.
Tools such as \lstinline{norm_num}, \lstinline{ring}, and \lstinline{norm_cast} arose after user requests
to automate mathematically trivial proofs.
Others, such as \lstinline{pi_instance},
make clever use of the metaprogramming API
to eliminate boilerplate code.
It is hard to separate the metaprogram components of \mathlib
from the mathematical formalizations,
as many tactics were inspired by particular patterns of use
and rely on nontrivial theories
or the proper use of attributes on library lemmas.


While our approach is not unique,
our emphasis is a shift from the traditional point of view:
we consider tactic and tool development as part of library design
rather than system development.
This division of labor has been possible thanks to the flexibility
of metaprogramming in Lean.


\section{Community}
\label{section:community}


The \mathlib library is not developed to support any single project.
The interests of its contributors range from research mathematics, to STEM education, to automated proof search, to program verification.
These contributors come primarily from academia; some are renowned researchers and some are bachelor students.
Design decisions and future directions are discussed openly and publicly, with little central control over the library's content.

That a cohesive library has been developed by a community with such diverse motivations and backgrounds is perhaps surprising.
Unlike most proof assistant libraries, which are developed or overseen by dedicated research groups, \mathlib is organized as an open-source community project.
While we have witnessed many of the familiar difficulties with such projects, this organizational scheme is arguably one of the reasons the library has been successful.

The communal nature of the project is, perhaps, one reason that Lean and \mathlib have attracted many mathematicians.
Mathematical topics are delicately intertwined.
Rather than individually building projects on top of a generic core library, the involved mathematicians have integrated their projects with each other and with \mathlib itself.
To many, the main research question in \mathlib is how to design a library of formal mathematics that is both broad and deep.
This question cannot be answered without the collaboration of many people; the range of contributors is a vital aspect of the project.

A side effect of the adoption of \mathlib by mathematicians is that Lean is being used in mathematics education.
While it is not uncommon to see proof assistants in certain computer science courses, we know of few cases where they have been introduced to undergraduate mathematicians~\cite{Neup19}.
Some of these undergraduates interact on the Lean Zulip chat room, contribute to \mathlib, and participate in the review process.

\subsection{GitHub and Zulip}
\label{subsection:community:github}

Communication about \mathlib occurs mainly over two channels.
The project is hosted on GitHub,\footnote{\url{https://github.com/leanprover-community/mathlib/}}
where contributions in the form of pull requests can be reviewed and edited.
More casual conversations occur in a Zulip chat room.\footnote{\url{https://leanprover.zulipchat.com/}}

Eleven community members have been designated as maintainers of the library.
These people are responsible for approving pull requests
in their areas of expertise.
Pull request reviews are welcomed from all members of the community;
in addition to the formal process on GitHub, PRs are often discussed on Zulip.
Community members are given write access to non-master branches of the repository,
and many PRs are developed as group efforts on these branches.

The Zulip chat room is home to a broad range of discussions.
In particular, a channel for new members welcomes elementary questions about Lean and formalization.
Another channel focuses on formalizing advanced mathematics.
A number of people have cited the accessibility of this chat room as a reason for deciding to use Lean.

\subsection{Projects Based on \mathlib}
\label{subsection:community:oneoff}

As well as being a cohesive collection of mathematics on its own, \mathlib should also serve as a library on which to build more specialized projects.
Users have undertaken many such projects, and the variety of topics reflects the diverse interests of the \mathlib community.
These projects range from published research papers to collaborative weekend efforts coordinated on Zulip.
We list here a non-exhaustive selection, to give an idea of what \mathlib can support.

\paragraph{The cap set problem} In 2016, Ellenberg and Gijswijt discovered a solution to the cap set problem, a longstanding open question in combinatorics.
Their celebrated proof~\cite{Elle17}, published in the \emph{Annals of Mathematics} in 2017, was noted for its use of elementary methods from linear algebra.
Dahmen, H\"olzl, and Lewis~\cite{Dahm19} formalized this result in 2019.
The formalization is based on \mathlib and resulted in significant contributions to the linear algebra theory (Section~\ref{section:linalg}), as well as those related to finite combinatorics.
It is a rare example of contemporary research mathematics being formalized in a proof assistant, made possible by collaboration between a mathematician and experts in formalization.

\paragraph{The continuum hypothesis} Han and van Doorn~\cite{Han19} verified the unprovability in ZFC of the continuum hypothesis.
They have since shown the unprovability of $\neg\text{CH}$~\cite{Han20}, completing the notoriously intricate full independence proof.
The formalization builds on the \mathlib embedding of ZFC and develops the syntactic theory of first order logic.
The independence of CH was one of the few remaining results
on Wiedijk's list of formalization targets\footnote{\url{http://www.cs.ru.nl/~freek/100/}}
that had yet to be formalized in any proof assistant.

\paragraph{Perfectoid spaces} Scholze was awarded a Fields Medal in 2018, in part for introducing the definition of a perfectoid space.
To test the ability of a proof assistant to understand extremely complicated mathematical structures, Buzzard, Commelin, and Massot defined perfectoid spaces in Lean~\cite{BCM20}.
Defining this structure relies on a large amount of algebraic and topological theory, and the months-long effort to formalize it resulted in thousands of lines of Lean code.

\paragraph{The sensitivity conjecture} Huang's sensational proof of the boolean sensitivity conjecture~\cite{Huan19} in 2019 was widely discussed, including on the Lean Zulip chat.
Following Knuth's simplified writeup,\footnote{\url{https://www.cs.stanford.edu/~knuth/papers/huang.pdf}} a number of community members formalized the proof in a matter of days.
The formalization reuses linear algebra machinery from the cap set project and amounts to fewer than 450 lines of heavily commented code.

\paragraph{Cubing a cube} After a challenge was issued at the 2019 Big Proofs meeting in Edinburgh, van Doorn formalized J.~E.\ Littlewood's ``elegant'' proof that any dissection of a cube into smaller cubes must contain at least two cubes of equal width.
This had been considered an exemplar of a proof with a simple intuitive argument that is hard to make formal.
It was another remaining item on Wiedijk's list of targets.

\paragraph{Decision procedures for modal logics} Wu and Gor\'e~\cite{Wu19} verified decision procedures for the modal logics K, KT, and S4. The decision procedures are formalized as programs in Lean with total correctness proved. The formalization makes extensive use of sublist permutation which is fully supported by \mathlib. The need for proving termination which involves reasoning about modal degrees has in turn resulted in further development of list theory of \mathlib.

\section{Comparison with Other Libraries}
\label{section:comparison}

In this section, we compare and contrast \mathlib with other substantial formal libraries for mathematics, including libraries for Mizar~\cite{Banc15, Grab10}, HOL Light~\cite{Harr09}, Isabelle/HOL~\cite{Nipk02}, Coq/SSReflect~\cite{Bert04,Mahb17}, and Metamath~\cite{Meg19}.
Our goal here is not to provide detailed comparisons of the various design choices, but, rather to sketch the design space in broad strokes, situate \mathlib within it, and explain some of the decisions we have made.
Our choice of comparisons is not meant to be exhaustive: there are also substantial mathematical libraries in HOL4~\cite{Slin08}, ACL2~\cite{Kau13}, PVS~\cite{Owre92}, and NuPRL~\cite{Const86}, as well as notable libraries built on standard Coq, such as the Coquelicot analysis library~\cite{Bold15}.

\subsection{The Design Space}
\label{subsection:design:space}

We will focus on three specific aspects of the design space, namely, the choice of axiomatic foundation, the mechanisms used to represent structures and the relationships between them, and the use of automation.

Most contemporary interactive proof systems are based on either set theory, simple type theory, or dependent type theory.
The decision to use an untyped framework like set theory or a typed alternative is fundamental.
Types play two important roles.
First, input in a typed system is often less verbose because users can elide details that can be inferred from the types of the objects involved.
For example, users can write \lstinline{x + y} and allow the system to infer the meaning of the plus sign from the types of its arguments.
Second, a type system can more easily catch and report errors, such as sending the wrong number or kinds of arguments to a function, or sending them in the wrong order. Another important choice is whether or not the axiomatic framework is constructive, and, in particular, whether it specifies a computational behavior for objects defined in the framework.

\subsection{Libraries Based on Set Theory}

As an axiomatic foundation, set theory has two important advantages.
First, it is accepted by many mathematicians as being the official foundation for mathematics, or, at least, an uncontroversial one.
And, second, it is fairly easy to implement and check proofs in this framework.

The Metamath system~\cite{Meg19} is a generic system for representing formal axiomatic frameworks, but its largest library is built on set theory.
That library currently has about 23,000 theorems and 600k lines of code, covering algebra, analysis, topology, number theory, and other areas.
Because the foundation is so simple, the source files need to provide explicit proofs that formulas are well formed, and there are front ends that insert that automatically.
But the system uses very little automation beyond that.
There are a number of reference checkers on offer that can check the entire library in a few seconds.
Sets are used to relativize quantifiers to specific domains.
For example, the following states that every continuous function on a closed interval is bounded:
\begin{verbatim}
  ((A ∈ ℝ ∧ B ∈ ℝ ∧ F ∈ ((A[,]B)–cn→ℂ)) →
    ∃x ∈ ℝ ∀y ∈ (A[,]B)(abs‘(𝐹‘y)) ≤ x)
\end{verbatim}
The need to write proofs that are fully detailed and explicit, however, places a high burden on the user.

Mizar~\cite{Banc15, Grab10}, which dates back to the early 1970s, is based on set theory with Grothendieck-Tarski universes.
Its vast library~\cite{Banc18} currently contains over 3.1 million lines of code and spans many fields of mathematics.
Proofs in the system are designed to mirror mathematical vernacular.
They are written in a declarative way, and a fixed checker determines whether inferences are valid~\cite{Grab10}.
To support a structure hierarchy, Mizar uses a \emph{soft typing} system whereby users register associations
that are used by the checker.
The library has developed quite an elaborate hierarchy in this way~\cite{Grab16} (see also~\cite{Grab10,Kali17}).
To our knowledge, there is no formal specification of the inferences that are accepted by the system,
making it hard to implement an independent reference checker.
Nonetheless, Kaliszyk et al.\ have had success reimplementing the system in the Isabelle framework and checking some of the Mizar files~\cite{Kali16, Kali17}, and Urban and Sutcliffe have shown that Mizar's inferences can be cross validated by automated theorem provers~\cite{Urb08}.

\subsection{Libraries Based on Simple Type Theory}

A number of contemporary proof systems implement versions of \emph{simple type theory}, among them HOL4~\cite{Slin08}, HOL Light~\cite{Harr09}, and Isabelle/HOL~\cite{Nipk02}.
In simple type theory, types and objects are separate things; one defines types and type constructions, and then one defines objects of those types.
Types cannot depend on objects: for example, one cannot define a type $\mathbb{R}^n$ that depends on a natural number parameter $n$.

Simple type theory excels at dealing with concrete structures like the integers, the reals, and the complex numbers,
but when it comes to algebraic reasoning, the fact that types cannot depend on parameters is a severe restriction.
Structures in mathematics are often parameterized by other objects: for $n \ge 1$, the type of $M_n(R)$ of $n \times n$ over a ring form a ring, and for every prime number $p$, the integers modulo $p$, $\mathbb{Z}/p\mathbb{Z}$, form a field; $L^p$ spaces, rings of $p$-adic integers, and spaces $C^k(X)$ consisting of $k$-times continuously differentiable real-valued functions on a subset $X$ of the reals all depend on parameters.
In simple type theory, one has to model these using predicates on fixed ambient types, erasing many of the benefits of using type theory in the first place.
Another limitation is that one cannot form spaces or structures whose elements are structures, such as the space of nonempty compact metric spaces with the Gromov-Hausdorff distance or the category of groups in some universe.

HOL Light~\cite{Harr09} is John Harrison's implementation of simple type theory, inspired by Mike Gordon's original HOL system.
The library, close to 800k lines of code, is especially strong in multivariate real analysis and complex analysis.
It served as the basis for the Flyspeck project~\cite{Hales15}.
Harrison uses a trick~\cite{Harr13} to formalize theorems about $\mathbb{R}^n$ for arbitrary $n$, using type variables to stand proxy for their cardinality.
But this trick has limited utility:
in simple type theory, one cannot even state that for every $n$ there is a type of cardinality $n$, and so there is no way of instantiating such a generic theorem to particular parameters.

Many aspects of our structure hierarchy, including the algebraic hierarchy, the axiomatization of topological spaces and normed spaces, our development of measure theory, and our use of filters in analysis, were modeled after similar developments in Isabelle~\cite{Nipk02}. Isabelle's extensive library has about 900k lines of code, and its Archive of Formal Proofs~\cite{Blan15} has an additional 2.3 million lines of code.
To treat common data types as instances of structures, Isabelle adopts a conservative extension of simple type theory with \emph{axiomatic type classes}~\cite{Haft06,Wenzel97}.
But type classes can only depend on a single type parameter, and, moreover,
the use of type classes suffers from the limitations imposed by simple types.
For example, the extensive libraries and automation developed for instances of rings like the integers and the reals cannot be applied to the ring of integers modulo some number $m$.
For such structures, Isabelle also provides the mechanism of \emph{locales}~\cite{Ball03,Kamm99}, but, once again, this means relinquishing some of the benefits of the use of type theory in the first place.
We therefore consider it a substantial accomplishment that we are beginning to approximate the depth and range of Isabelle's structure hierarchy in a dependently typed setting.

Isabelle's supporting automation is especially good.
The system provides internal automation, decision procedures, and a conditional term rewriter~\cite{Nipk02},
but can also call external resolution provers and SMT solvers and reconstruct their proofs~\cite{Blan13,Blan16}.
Using axiomatic type classes, internal automation can work generically with types that instantiate the relevant structures, just as our \lstinline{norm_num} works for types that support numerals and the relevant operations.
Isabelle's automation is much more powerful than that of Lean and \mathlib, but we again consider it notable that we are making progress towards the use of such automation with dependent types.

\subsection{Libraries Based on Dependent Type Theory}
\label{subsection:comparison:mathcomp}

Turning to dependent type theory, the best point of comparison for \mathlib is the \mathcomp library~\cite{Mahb17}, based on Coq and the SSReflect proof language~\cite{Gon10}.
Coq's version of dependent type theory is very similar to Lean's.
The \mathcomp library was the basis for the landmark formalization of the odd order theorem~\cite{Gon10},
and focuses on related parts of mathematics, including group theory, linear algebra, and representation theory.
The library itself is about 110k lines of code, not including results specific to the odd order theorem. Other libraries have been built on it, including an analysis library~\cite{Aff18} that is still under development.

In contrast to \mathlib, the \mathcomp library is constructive throughout (although the analysis library just mentioned uses classical logic).
Another distinguishing feature of the \mathcomp library is that it uses very little external automation, and instead relies heavily on the computational interpretation of the underlying axiomatic framework, so that inferences can be verified by computational unfolding of expressions.

Like \mathlib, \mathcomp relies on powerful elaboration mechanisms to support a structure hierarchy, though the specific mechanisms are different: instead of type classes, \mathcomp uses \emph{canonical structures}~\cite{Garr09}, a particular kind of unification hint~\cite{Asp09}.
The core algebraic hierarchy in the \mathcomp library~\cite[Section 7.3]{Mahb17} includes key structures such as $\mathbb{Z}$-modules (essentially, abelian groups), rings, commutative rings, fields, algebraically closed fields, modules, and algebras.
Other parts of the library define vector spaces, and structures with orders and norms, that are designed to support reasoning about subfields of the complex numbers.
There are additional structures to support group theory and representation theory,
and the analysis library includes structures such as topological spaces and normed modules.

The \mathcomp library is generally more conservative than we are when it comes to instantiating structures and substructures.
For example, the theory of the natural numbers in \mathcomp is largely self contained, whereas our natural numbers instantiate an ordered semiring, which in turn inherits from the structures of additive and multiplicative monoids, linear orders, and so on. A calculation involving the natural numbers in \mathlib may, and usually does, involve generic facts from all levels of this hierarchy.

The \mathcomp library is also carefully designed to avoid the need for the kinds of searches described in Section~\ref{section:typeclasses}.
An artifact of the use of canonical structures is that concrete structures have to be instantiated to all the classes they inherit from.
For example, the integers are declared as a $\mathbb{Z}$-module, a ring, a commutative ring, an integral domain, and so on.
In \mathlib, declaring an instance of a structure automatically handles not just substructures, but also induced structure.
When we declare the reals to be a metric space, for example, it thereby inherits the structure of a uniform space hence a topological space.

\section{Conclusion}
\label{section:conclusion}

With respect to the design space described in the previous section, the characteristic features of \mathlib are as follows.
First, it is based on a dependent type theory.
We have chosen a typed framework for the reasons indicated in Section~\ref{subsection:design:space}, and given that we want to carry out the full range of mathematical constructions, judicious use of dependent types is unavoidable.
Second, it is focused on contemporary mathematics, which is resolutely classical.
Nonetheless, large portions of the library are explicitly computational, and functions, say, on lists and integers can be executed.
Third, it incorporates useful automation, such as a good conditional simplifier and domain-specific tactics written in the system's metaprogramming language.
Our automation is still in an incipient state, but we hope and expect to expand these capabilities.
Finally, it includes a large and interconnected hierarchy of mathematical structures and instances.
Each of the other libraries discussed in Section~\ref{section:comparison} shares some of these features, but we know of no other library that shares them all.

\begin{figure}
 \includegraphics[width=\columnwidth]{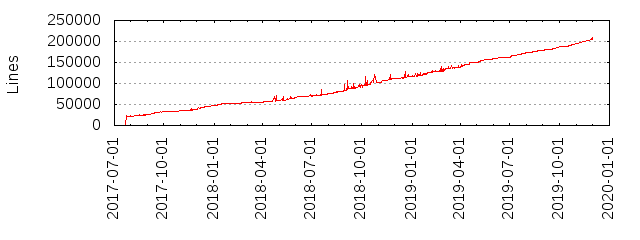}
 \caption{Lines of code in the \mathlib repository over time, including white space, comments, and auxiliary files.}
 \label{figure:loc_graph}
\end{figure}

\begin{figure}
 \includegraphics[width=\columnwidth]{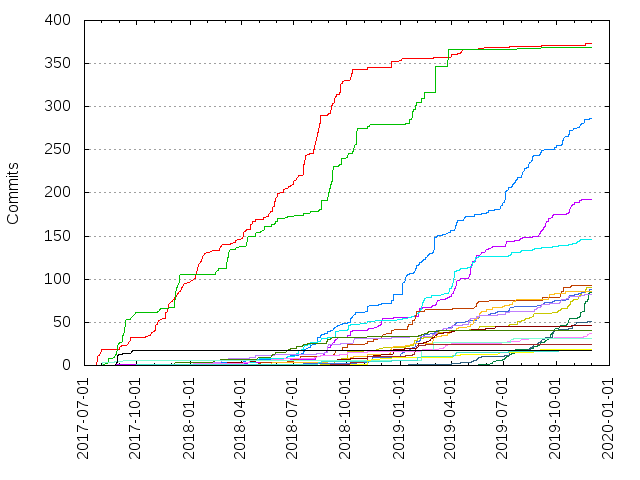}
 \caption{Number of commits to the \mathlib repository over time, by the 20 most active authors.}
 \label{figure:commits}
\end{figure}

The \mathlib library and its surrounding community continue to grow (Figures~\ref{figure:loc_graph} and~\ref{figure:commits}).
Various individuals and groups are actively working to expand its mathematical content,
automated reasoning tools,
and surrounding infrastructure.
There is no target or end goal in mind:
as in traditional mathematics, the development of new theories will follow the needs and desires of the people involved.

A new version of Lean is currently in development~\cite{Ullr19}.
It will not be backward compatible with the current version,
and when the system is ready, we expect to update \mathlib.
At the current time, it is hard to predict the scale of this undertaking.
Lean 4 promises even more efficient metaprogramming
with access to input syntax trees and a customizable parser;
it might be possible to automate much of the porting process.

The design of \mathlib builds on the lessons of existing libraries,
but the library's special character can be attributed to
the range 
of its contributors.
The combined efforts of mathematicians, computer scientists, and programmers
to design a single unified library of formalizations and tools
has
been successful,
and we are still learning from one another,
as the community continues to evolve.

\begin{acks}                            

We thank Jasmin Blanchette, Assia Mahboubi, Josef Urban, and the anonymous referees for their helpful comments and suggestions.  
  
The following people have either authored commits in the \mathlib repository or are attributed authors of \mathlib files:

Lucas Allen, Ellen Arlt, Jeremy Avigad, Tim Baanen, Seul Baek, Reid Barton, Tim Baumann, Alexander Bentkamp, Alex Best, Aaron Bryce, Kevin Buzzard, Louis Carlin, Mario Carneiro, Bryan Gin-ge Chen, Johan Commelin, Sander Dahmen, Floris van Doorn, Gabriel Ebner, Ramon Fern\'andez Mir, Fabian Gl\"ockle, S\'ebastien Gou\"ezel, Tobias Grosser, Jesse Han, Keeley Hoek, Johannes H\"olzl, Michael Howes, Simon Hudon, Christopher Hughes, Michael Jendrusch, Sangwoo Jo, Kevin Kap\-pel\-mann, Parikshit Khanna, Yury Kudryashov, Joey van Langen, Kenny Lau, Sean Leather, Guy Leroy, Robert Y.\ Lewis, Amelia Livingston, Isabel Longbottom, Jean Lo, Paul-\-Nico\-las Madelaine, Patrick Massot, Bhavik Mehta, Rohan Mitta, Stephen Morgan, Scott Morrison, Leonardo de Moura, Oliver Nash, Wojciech Nawrocki, Moreni\-keji Neri, Casper Putz, Matt Rice, Mitchell Rowett, Jan-David Salchow, Blair Shi, Rory Qianyi Shu, Calle S\"onne, Robert Spencer, Neil Strickland, Abhimanyu Pallavi Sudhir, Callum Sutton, Andreas Swerdlow, Nathaniel Thomas, Alistair Tucker, Sebastian Ullrich, Koun\-din\-ya Vajjha, Jens Wagemaker, Minchao Wu, Hai\-tao Zhang, Zhou\-hang Zhou, Sebastian Zimmer, Andrew Zipperer.

\end{acks}

\bibliography{mathlib-paper}

\end{document}